\documentstyle[12pt]{article}
\title{
Poisson Brackets Scheme for Vortex Dynamics in Superfluids and
Superconductors and Effect of Band Structure of Crystal.}
\author{ G.E. Volovik\\
Low Temperature Laboratory, Helsinki University of Technology\\
Otakaari 3A, 02150 Espoo, Finland\\
and\\
L.D. Landau Institute for Theoretical Physics, \\
Kosygin Str. 2, 117940 Moscow, Russia\\
}
\begin{document}
\maketitle
\begin{abstract}
{Poisson brackets for the Hamiltonian dynamics of vortices
are discussed for 3 regimes, in which the dissipation can be
neglected and the vortex dynamics is reversible:  (i) The
superclean regime when the spectral flow is suppressed.  (ii) The
regime when the fermions are pinned by crystal lattice. This
includes also the regime of the extreme spectral flow of fermions
in the vortex core: these fermions are effectively pinned by the
normal component. (iii) The case when the vortices are
strongly pinned by the normal component. All these limits are
described by the single parameter
$C_0$, which physical meaning is discussed for superconductors
containing several bands of electrons and holes. The effect of the
Fermi-surface topology on the vortex dynamics is also discussed. }
\end{abstract}

{\it 1. Introduction.}
 The problem of the vortex motion is well understood for the
translational invariant Fermi superfluids
\cite{Kopnin-Kravtsov,KopninLopatin1995,KopninVolovik,Stone}.
When the theory is applied to superconductors, one should take
into account the effect of the band structure of the crystal on
the vortex dynamics. In this case the topology of the Fermi-surface
is to be important and we make an attempt to consider this effect in
the regimes, when the dissipation is small and can be
neglected.

{\it 2. Poisson Brackets Formalism for the Vortex Motion.}
 We start with the phenomenological
hydrodynamic  equations for the system of distributed
vortices in superfluids or superconductors.  In the limit of
vanishing dissipation the dynamics of the collective variables
becomes conservative and in principle can be described by the
effective action. However, as usually occurs in the hydrodynamic
systems, such action is not well defined and the Hamiltonian
formalism in terms of the Poisson brackets (PB) becomes preferrable.
In this formalism the Hamiltonian is the function of the relevant
hydrodynamic variables characterized by the algebra of
the Poisson brackets.

We assume that the normal component is
clamped, ie its velocity ${\bf v}_n=0$. This is typical for
superfluid $^3$He due to its high viscosity and for superconductors
where ${\bf v}_n$ is fixed by the impurities in the crystal
lattice. The remaining hydrodynamic variables at low temperature
are the  mass density
$\rho$ and the superfluid velocity ${\bf v}_s$, which is
non-potential in the presence of the distributed vorticity. The
Hamiltonian
$$H=\int d^3r~ [{1\over 2}\rho_s^{ij} v_s^i v_s^j + \epsilon
(\rho)]~~,\eqno(2.1)$$
contains the internal energy density $\epsilon
(\rho)$ and the kinetic energy.
In crystals, due to absence of the Galilean invariance, the
superfluid  component does not coincide with the  density
$\rho$ of the electrons even at $T=0$: $\rho_s(T=0)\neq
\rho$.  Actually  $\rho_s$ can be much
smaller than $ \rho$ because most of the electrons are concentrated
in the completely filled bands. Nevertheless the hydrodynamic
equations are valid for the descrition of the long-wave-length
dynamics of
$\rho$ and ${\bf v}_s$.

The motion equations are obtained as the Liouville equations
$${\partial \rho\over \partial t} =\{H,\rho \}~~,~~{\partial {\bf
v}_s\over
\partial t} = \{H,{\bf v}_s \} ~~.\eqno(2.2)$$
if one uses the PB between the variables. These PB
are universal, ie they do not depend on the
Hamiltonian
\cite{DzyaloshinskiiVolovick}, and we propose:
$$\{\rho({\bf r}),{\bf v}_s({\bf r}')\}= \vec\nabla
 \delta( {\bf r}-{\bf
r}')~~,\eqno(2.3)$$
$$\{\rho({\bf r}),\rho({\bf r}')\}=0~~,\eqno(2.4)$$
$$\{ v_{si}({\bf r}),v_{sj}({\bf r}')\}=-e_{ijk}
{(\vec\nabla\times {\bf
v}_s)_k
\over \rho -C_0}\delta( {\bf r}-{\bf r}')~~.\eqno(2.5)$$
The first two PB are conventional (see
\cite{DzyaloshinskiiVolovick}), eg the
Eq.(2.3) follows from the fact that the particle number
and the phase of the
condensate are canonically conjugated variables.   The Eq.(2.5)
contains a new variable $C_0$, which is to be the dynamical
invariant of the system,
${\partial C_0\over\partial t}=0$. With this constraint the Poisson
brackets satisfy the Jacobi identity
$\{a\{bc\}\}+\{b\{ca\}\}+\{c\{ab\}\}=0$. The Eq.(2.5) with $C_0=0$
was derived in superfluids in the $T=0$ limit
\cite{VolovikDotsenko}. In derivation it was assumed that each
element of the vortex moves with the local superfluid velocity
${\bf v}_s ({\bf r})$, which corresponds to the Helmholtz theorem
for the perfect liquid. The PB Eq.(2.5) with $C_0=\infty$ was
written in
\cite{DzyaloshinskiiVolovick}. This corresponds to the motion of
vortices with the local normal velocity ${\bf v}_n ({\bf r})$. This
is actually a true hydrodynamic regime, which requires the complete
pinning of the vortex lines by the normal component or by the heat
bath of the crystal lattice
\cite{AndreevKagan}.

The range  $0<C_0<\infty$ corresponds to the
intermediate regimes when the vortex is unpinned, but some
groups of particles (electrons) are pinned by the heat bath
during the vortex motion. In addition to the trivial
localization of the electrons by the crystal lattice and
impurities, the quasiparticles are also pinned by the anomalous
process of the spectral flow in the core of the vortex, discussed in
\cite{Q-modes-Index,ZhETF,KopninVolovik,Stone}. When the vortex
moves in the regime of the extreme spectral flow, the momentum is
effectively  transferred  from the vortex to the heat bath, which
corresponds to the pinning of fermions by the heat bath.

With the PB in Eqs.(2.3-5), the Liouville equations (2.2) of the
nondissipative dynamics become
$${\partial \rho\over \partial t} + \vec\nabla\cdot (\hat\rho_s
{\bf v}_s)=0 ~~,\eqno(2.6)$$
$${\partial {\bf v}_s\over \partial t} +{1\over \rho
-C_0}(\hat\rho_s{\bf v}_s)\times(\vec\nabla\times {\bf
v}_s)+\vec\nabla\mu=0~~,~~\mu ={\delta H\over \delta\rho}
~~.\eqno(2.7)$$

Using the kinematic definition of local
velocity ${\bf v}_L$ of vortex
lines\cite{SoninReview,DzyaloshinskiiVolovick}:
$$\partial_t {\bf v}_s +\vec\nabla\mu=-
{\bf v}_L\times(\vec\nabla\times {\bf
v}_s) ~~,\eqno(2.8)$$
one obtains the following relation between ${\bf v}_L$ and ${\bf
v}_s$:
$$(\rho -C_0){\bf v}_L= \hat\rho_s {\bf v}_s~~.\eqno(2.9)$$
For superconductors the lhs of Eq.(2.9) determines the
Hall conductivity in the limit of vanishing dissipation:
$$\sigma_{\rm Hall}={ec\over m B}(\rho -C_0)~~,$$
where $B$ is magnetic field.

The equation (2.9) is not Galilean invariant. The Galilean
invariance is restored by introducing the velocity of the normal
component
${\bf v}_n$, which coincides with the velocity of crystal lattice
in the case of superconductors. Then multiplying by the circulation
$\vec\kappa$ (with
$\vert
\vec\kappa\vert =N\pi \hbar/m$, and $N$ being the winding number of
the vortex)  one obtains the equation for the balance of Magnus,
spectral flow and Iordanskii forces acting on the vortex
\cite{KopninVolovik}:
$$\vec\kappa \times [\rho ({\bf v}_s-{\bf v}_L) + C_0({\bf
v}_L-{\bf v}_n) +
\hat\rho_n ({\bf v}_n-{\bf v}_s)]=0~~,\eqno(2.10)$$
where $\hat\rho_n=\rho -\hat\rho_s$ is the tensor of the normal
density.

The density $\rho$ and the superfluid density
tensor $\hat\rho_s$ are well determined quantity, the latter being
determined by the current-current correlation function. Let us now
discuss the parameter $C_0$. For the systems  with
translational invariance \cite{KopninVolovik,Q-modes-Index} the
parameter $C_0$  was determined only in the special
limit case, the so called hydrodynamic regime. As we see
below, in both hydrodynamic and collisionless regimes the
dissipation can be neglected and thus the  Hamiltonian approach is
valid. As a result the Eqs.(2.9-10) are valid in both
regimes, but with different values of the parameter
$C_0$. For example, as follows from Refs.
\cite{KopninVolovik,Q-modes-Index}, in the  systems  with the
translational invariance the collisionless value of the parameter
$C_0$ in Eqs.(2.9-10) is zero. But for
superconductors with several electronic bands the situation is more
complicated and the parameter $C_0$ can be nonzero even in the
collisionless regime. Moreover one can have the
collisionless regime for one band and the hydrodynamic regime for
the other band.

{\it 3. Particles and holes contributions.}
The low-frequency dynamics and thermodynamics of the
fermi-liquid or superconductors are determined by the low-energy
quasiparticles. In the same way the low-frequency dynamics of
vortices is determined by the low-energy excitations in the vortex
core. The latter are concentrated on the anomalous branch of the
spectrum \cite{Caroli}. Here we follow the simplified version of
\cite{Kopnin-Kravtsov,KopninLopatin1995}  (see
Refs.\cite{Q-modes-Index,Stone}). Let us start with the
axisymmetric vortex in the translational invariant surrounding.
The  spectrum of the low-energy excitations in the core is  defined
in the frame of the moving vortex, where the Hamiltonian
does not depend on time, if there are impurities. The energy
of the low-energy branch is expressed in terms of the canonically
conjugated variables: the angular momentum
$Q$ and the angle $\alpha$ of the linear momentum, ${\bf
k}=(k_F\cos\theta, k_F\sin\theta
\cos\alpha, k_F\sin\theta\sin\alpha)$, in the transverse plane:
$$
{\cal H}= C\omega_0 Q + {\bf v}_s\cdot{\bf k}~~.
\eqno (3.1)
$$
The first term describes the orbital motion of fermions bound
in the core around the vortex axis \cite{Caroli}.
The effect of the vortex on the motion of the quasiparticle is
similar to the magnetic field. The quantization of this orbital
motion leads to the descrete levels (see also Sec.4) with either
integer or half-odd integer generalized angular momentum $Q$
\cite{MissirpashaevVolovik1995}. The frequency of rotation
$\omega_0$ is the function of $\theta$. The direction of rotation
is determined by the "chirality" factor $C=\pm 1$. For the
conventional case of the particle-like excitations in the vortex
with winding number $N=1$ one has
$C=-1$ \cite{Caroli}, while $C=1$ for the hole-like excitations in
the vortex with the same winding number. This follows from the
index theorem, which relates the number of fermion zero modes in the
vortex core to the vortex winding number
$N$ \cite{Q-modes-Index}: in the case of holes the
topological invariant, which determines the
number of zero modes, changes sign.

The second term is the
energy due to the superflow in the vortex frame.

When the vortex moves with respect to the heat bath, its dynamics
is nonequilibrium. The kinetics of the fermions on the branch in
Eq.(3.1) is governed by the Boltzmann equation for the distribution
function, $n(Q,\alpha)$, in the $Q-\alpha$ phase space\cite{Stone}:
$$
{\partial  n\over \partial t}+C\omega_0 {\partial n\over
\partial \alpha} -{\partial  ( ({\bf v}_s - {\bf
v}_L)\cdot{\bf k}) \over
\partial \alpha}  ~{\partial n\over
\partial Q}= -{n(Q,\alpha)-  n_{\rm
eq}(Q,\alpha)\over \tau}
\eqno (3.2)
$$
The last term describes the relaxation to the equilibrium
distribution function $ n_{\rm eq}$ determined by the heat bath
outside the core:
$$ n_{\rm eq}(Q,\alpha)= f({\cal H}- {\bf v}_n\cdot{\bf k})=
f(C\omega_0 Q + ({\bf v}_s - {\bf v}_n)\cdot{\bf k})~~, \eqno
(3.3)$$
where $f(E)=(1+\exp(E/T))^{-1}$ is the Fermi-function.

If $\tau$
does not depend on $Q$ one gets the equation for average
momentum
$$\partial_t {\bar{\bf k}}+C\omega_0 {\hat {\bf z}}\times {\bar{\bf
k}} +{C\over 4}k_F^2\sin^2\theta
{\hat {\bf z}}\times ({\bf v}_n - {\bf v}_L)
(f(\Delta(T))-f(-\Delta(T)))= - {{\bar{\bf k}}\over
\tau} ~,
\eqno (3.4)
$$
$$ {\bar{\bf k}}=  {1\over 2}\int dQ {d\alpha\over 2\pi}
(n(l,\alpha)- n_{\rm
eq}(Q,\alpha)) {\bf k}~~.
\eqno (3.5)
$$

Here we take into account that $\int dQ \partial_Q n$ is limited by
the bound states below the gap $\Delta(T)$, since above the gap
$\Delta(T)$ the spectrum of fermions is continuous
\cite{KopninLopatin1995}. The effective interlevel distance for the
unbound (delocalized) states is $\omega_0=0$ and they will be
considered below.

In the steady state of the vortex motion one has $\partial_t
{\bar{\bf k}}=0$ and  the Eq.(3.4) is easily solved. The rhs of
Eq.(3.4) gives the momentum flow to the heat bath and thus  the
following force due to bound states below
$\Delta(T)$
$${\bf F}_{loc}=\int {dk_z\over 8\pi}  k_F^2\sin^2\theta
{1\over 1+  \omega_0^2\tau^2}
\tanh {\Delta(T)\over 2T} [({\bf v}_L - {\bf v}_n)
\omega_0\tau -C{\hat {\bf z}}\times({\bf v}_L - {\bf v}_n)]  ~.
\eqno (3.6)$$
The spectral flow of unbound states above $\Delta(T)$ is not
suppressed, since the  corresponding $\omega_0\tau=0$. This gives
$${\bf F}_{deloc}=-C{k_F^3\over 8\pi}\int  d\cos\theta
~ \sin^2\theta  \left(1- \tanh
{\Delta(T)\over 2T}\right) {\hat {\bf z}}\times({\bf v}_L - {\bf
v}_n)  ~.
\eqno (3.7)$$
Thus the total nondissipative spectral-flow force is
$${\bf F}_{nondiss~sp~flow}=-C{k_F^3\over 8\pi}\int  d\cos\theta
~ \sin^2\theta \left[1 - {\omega_0^2\tau^2 \over 1+
\omega_0^2\tau^2}\tanh {\Delta(T)\over 2T} \right]{\hat {\bf
z}}\times({\bf v}_L - {\bf v}_n)  ~.
\eqno (3.8)$$

The dissipative part of the spectral-flow force
$${\bf F}_{diss~sp~flow}=({\bf v}_n - {\bf v}_L){k_F^3\over
8\pi}\int  d\cos\theta ~ \sin^2\theta  \tanh {\Delta(T)\over
2T}~{\omega_0\tau
\over 1+
\omega_0^2\tau^2}
  ~,
\eqno (3.9)$$
can be neglected  when
$\lambda=\omega_{0}\tau/(1+\omega_{0}^2\tau^2) \ll 1$. So, the
vortex motion is governed by the conservative Hamiltonian dynamics
either in the so called hydrodynamic regime, when
$\omega_{0}\tau  \ll 1$, or in collisional regime, when
$\omega_{0}\tau  \gg 1$. In both cases the contribution of the
spectral flow to the parameter $C_0$ in Eq.(2.10)
from one spin
direction acquires universal values at low $T$:
$${C_0\over m}=0~~,~~\omega_{0}\tau  \gg 1 ~~,\eqno(3.10)$$
$${C_0\over m}=V^p~~,~~\omega_{0}\tau  \ll 1~~,~~{\rm
particle~states} ~~,\eqno(3.11)$$
$${C_0\over m}=-V^h~~,~~\omega_{0}\tau  \ll 1~~,~~{\rm
hole~states} ~~.\eqno(3.12)$$
Here $V^p= p_F^3/6\pi^2$ is the volume of the Fermi-sphere of
particle states (with one spin direction), while $V^h= p_F^3/6\pi^2$
is the volume of the Fermi-sphere of the hole states.

In conventional superconductors there are no zeroes of energy in
the quasiparticle spectrum and thus the Fermi surface is absent.
Zeroes however  appear due to vortices. In the continuous
description of the vortex core, the zeroes in the classical energy
spectrum $E=\sqrt {\epsilon^2({\bf k}) +
\vert \Delta({\bf k},{\bf r})\vert^2}$ are concentrated on the
vortex axis, but can be splitted into point nodes distributed in the
vortex core
\cite{VolovikMineev1982}. The function ${\bf k}({\bf r})$, which
shows the position of zero of $E$ in the momentum space as a
function of the coordinate
${\bf r}$ in the real space, maps the cross section of the vortex
into the momentum space. Thus, if one sweeps the cross section of
the  $N$-quantum vortex, one obtains the closed surface in the
momentum space, swept by zeroes. The volume within this surface is
just $C_0/m$ times $N$ \cite{ZhETF}. The physical meaning of
$C_0/m$ is the number of the electronic states which remain kept
by the heat bath during the vortex motion. In the frame moving with
the vortex this corresponds to the number of the states flowing
from the vortex to the heat bath in the extreme spectral flow
regime.

{\it 4. Electronic bands in crystals, open orbits.}
Let us consider the crystal with anisotropic Fermi-surface. In the
case of the particle states the Eq.(3.11) remains to be valid for
the arbitrary closed surface of zeroes with $V^p$
being the volume within the surface \cite{ZhETF}. If one
neglects the dependence of $\omega_0\tau$ on $\alpha$, $\theta$
and $Q$ one can write an interpolating equation for the closed
surface of particle states:
$${C_0({\rm particles})\over  m}= { V^p \over
1+\omega_0^2\tau^2} .\eqno(4.1)$$
The contribution of the holes to $C_0$ is
$${C_0({\rm holes})\over  m}= V^B - V^h {1 \over
1+\omega_0^2\tau^2} =  V^B{\omega_0^2\tau^2 \over
1+\omega_0^2\tau^2} + V^p {1 \over
1+\omega_0^2\tau^2} ~~,\eqno(4.2)$$
where $V^B$ is the total volume of the Brillouin zone,
$V^h=V^B-V^p$ is  the volume of the hole states. The
Eq.(4.2) follows from two arguments. (1) The spectral flow
parameter $C_0$ changes sign for holes, as was discussed in the
previous Section. (2) The completely filled band should be
considered as uneffected by the motion of the vortex, the electrons
on these bands are completely pinned by the crystal lattice. This
corresponds to the limit of the extremely fast relaxation and thus
to the extreme spectral flow.

The Eq.(4.2) agrees with the microscopic calculations in
Ref.\cite{KopninLopatin1995}. If the gap $\Delta$ is small
compared to the Fermi energy, then
$\rho$ is very close to the volume of
all the particle states:  $\rho
\approx m\sum V^p$. If there is only one band and this band
contains holes, then  $\rho
\approx m(V^B-V^h)$. In the superclean regime, $\omega_0 \tau \gg
1$, the Eq.(4.2) gives
$\rho - C_0\approx m(V^B-V^h)-mV^B=
 -mV^h$. As a result $\sigma_{\rm
Hall}=-(ec/B)V^h$ in agreement with
Ref.\cite{KopninLopatin1995}.

The  Eqs.(4.1) and (4.2) should transform into each other during
the filling of the Brillouin zone, when the  Fermi
surface of particles transforms to the Fermi surface of holes.
However Eqs.(4.1) and (4.2) differ by the value $V^B
\omega_0^2\tau^2/( 1+\omega_0^2\tau^2)$. The key to the problem of
matching these  equations in the Lifshitz transition  is
provided by open orbits, which appear as intermediate stage
between particle and hole Fermi-surfaces. The open Fermi-surfaces
with complicated topology were discussed in the relation to the
Hall effect in the normal metal (see recent paper
\cite{NovikovMaltsev} and references there). Here we show that
in the presence of the open surfaces of zeroes the parameter
$\omega_0\tau$ is small and Eqs.(4.1) and (4.2) match each other.

In the  semiclassical approach
the energy of fermions in the vortex core:
$$E^2=\epsilon^2({\bf k}) +
\vert \Delta({\bf r})\vert^2~~.   \eqno(4.3)$$
The lowest energy levels  are concentrated in the vicinity of zeroes
${\bf k}_0$ of the spectrum $\epsilon({\bf k})$, ie close to the
former Fermi-surface  of the normal metal. Near
the Fermi-surface one has $\epsilon({\bf k})={\bf v}({\bf k}_0)\cdot
({\bf k}-{\bf k}_0)= -iv_\perp \partial_s$, where $s$ is the
coordinate along
${\bf v}_\perp={\bf v}({\bf k}_0)-{\hat{\bf z}}({\hat{\bf z}}\cdot
{\bf v}({\bf k}_0))$.  Close to the vortex axis one has
$$\vert \Delta({\bf
r})\vert^2\approx \gamma^2r^2=\gamma^2(s^2+b^2)~~,
\eqno(4.4)$$
where $b$ is the impact parameter. We assume for simplicity that
the core radius is large compared to the size of the electronic
orbits: since the result is of the topological origin, it should
not depend on the model. First we quantize the fast motion along
$s$. According to supersymmetry the lowest energy level of this
motion lies exactly at zero energy
\cite{Q-modes-Index}. As  a result the spectrum of the excitations
in the core is determined by the slow motion along $b$, ie along the
line of the intersection of the  Fermi-surface
$\epsilon({\bf k})=0$ with the plane $k_z=const$. It is given by
$E(k_\parallel,b)=\gamma b$, where $k_\parallel$ is the coordinate
along the line of zeroes in the momentum space. The quantization
of the slow motion,
$\oint dk_\parallel b(k_\parallel)=2\pi n$, with
 $b(k_\parallel)=E/\gamma$ gives the levels of bound states in the
core
$$E_n=-n\omega_0~~,~~{1\over \omega_0}={1\over \gamma} \oint
{dk_\parallel\over 2\pi}~~.  \eqno(4.5)$$
For the closed spherical Fermi-surface this leads to the
conventional result for the states in the
vortex with large core radius:
$$E_n=-n\omega_0~~,~~\omega_0={\gamma
\over k_F
\sin\theta} ~~.  \eqno(4.6)$$
For the open orbits the integral in Eq.(4.5) diverges which gives
$\omega_0=0$ as it was expected.

{\it 5. Discussion.}
The Eqs.(4.1) and (4.2) can be generalized to the case of
several bands. If there are no open surfaces of
zeroes, the total $C_0/m$ contains the positive contribution
from the particles, the negative contribution from
 holes and the positive contribution  $kV^B$ where $k$ is
the number of Brillouin zones which are either completely filled or
contain the hole states:
$${C_0\over  m}=\sum_a   V^p_a {1 \over
1+\omega_{0a}^2\tau_a^2}-\sum_b   V^h_b {1 \over
1+\omega_{0b}^2\tau_b^2} + k V^B~.\eqno(5.1)$$
This also includes the summation over the spin indices.

Each Fermi surface has its own $\omega_{0}\tau$. The conservative
Hamiltonian approach applies and the Eq.(5.1) holds, only if for
each band the
parameter $\lambda=\omega_{0}\tau/(1+\omega_{0}^2\tau^2)
\ll 1$.  If $\lambda$ are not small, the Eq.(5.1) can be considered
only as interpolation since the real $\omega_0$ and $\tau$ are
complicated functions of the impact parameter and momentum $p_z$.
The limiting cases, when the Eq.(5.1) holds, can include the cases
when $\omega_{0}\tau$ is small in one zone and large in the other.
Thus $C_0$ as a function of external parameters (doping or direction
of magnetic field) should have plateaus interrupted by regions where
one of the parameters $\lambda$ changes between 0 and 1.  The latter
occurs also during the change of the topology of orbits.

The Eq.(2.9) can be also applied to other inhomegeneous systems
which become homogeneous on a large scale, such as  Josephson
junction arrays (JJA). In some cases the corresponding quantities
$\rho$,
$\rho_s$ and $C_0$ can be obtained after averaging over the scale of
the inhomogeneity. In the system of the SNS contacts, the parameter
$\rho - C_0$ is again small in the hydrodynamic limit due to
approximate particle-hole symmetry \cite{MakhlinVolovik1995}. This
leads to the almost ballistic motion of vortices in the absence of
the supercurrent, when ${\bf j}_s=\hat\rho_s {\bf v}_s=0$. It is
still unclear whether the approximate cancellation of $\rho$ and
$C_0$ occurs in SIS contacts.

I thank N.B. Kopnin for illuminating discussions.
This work was supported through the ROTA co-operation plan of the
Finnish  Academy and the Russian Academy of Sciences and by the
Russian Foundation for Fundamental Sciences,  Grant No.
96-02-16072.

\end{document}